\documentclass[11pt,a4paper]{article}
\usepackage[T1]{fontenc}
\usepackage[utf8]{inputenc}
\usepackage{mathptmx}
\usepackage{textcomp}
\usepackage{amsmath,amssymb}
\usepackage[margin=2.5cm]{geometry}
\usepackage{longtable,booktabs,array,calc}
\usepackage[authoryear,round]{natbib}
\usepackage[hidelinks]{hyperref}
\usepackage{microtype}
\providecommand{\tightlist}{\setlength{\itemsep}{0pt}\setlength{\parskip}{0pt}}
\title{Trend-concordance grey incidence with likelinesses: closed-form small-sample inference without elicited prior densities}
\author{Fernando A. Otero\\[2pt]
\small Departamento de Matem\'atica, Facultad de Ingenier\'ia,\\
\small Universidad Nacional de Mar del Plata, Mar del Plata, Argentina}
\date{}
\begin{document}
\maketitle
\begin{abstract}
Grey incidence analysis ranks the factors of a system from short sequences without assuming a probability model, but the grade it returns has no sampling theory: nothing indicates whether an ordering reflects the system or the few data at hand. This paper reduces trend agreement between two sequences to the signs of their centred increments and analyses the resulting counts with Lovegrove's likelinesses, averages of probabilities over an underlying set of distributions that replace the elicitation of a prior density by a statement of what is possible. The result is a concordance incidence grade with closed-form inference: estimation, the likeliness that one factor outranks another, prediction of future concordance and a baseline for searches over many candidates, with bounded underlying sets encoding partial knowledge. Comparisons whose sign falls below the resolution of the data are treated as indeterminate; discarding them is exact when they carry no information, and otherwise the grade becomes a grey number whose width measures what the data cannot decide. The grade is symmetric in any system and resolves the conflict between the totality and pair-symmetry axioms by moving system dependence into inference. Simulations show that plug-in prediction underestimates variance by a factor of up to $1+m/n$ regardless of sample size, and that a fair-coin baseline yields up to 63\% spurious patterns in searches, where a bounded set keeps them below 5\%. Worked textbook examples and a regularisation benchmark with known truth illustrate the method.

\end{abstract}

\hypertarget{introduction}{%
\section{Introduction}\label{introduction}}

Grey incidence analysis ranks the factors of a system by how closely their sequences follow a reference sequence, and it is designed for the situation statistics handles worst: few observations and no defensible probability model \citep{liu2017}. Its output, however, is a single number per factor. Nothing in the method says whether the ordering of two grades reflects the system or the particular handful of observations at hand, nor how the grade should behave as observations accumulate.

The literature has documented the symptoms. Different incidence models frequently produce incompatible orderings of the same factors \citep{dong2018}; almost no model attains a zero lower bound, and the totality and pair-symmetry axioms cannot in general hold together \citep{xie2009}. Attempts to attach uncertainty to the grade have borrowed tools that assume what grey analysis declines to assume: error propagation needs a noise model \citep{saavedra2013}, and bootstrap resampling is fragile at the sample sizes for which grey methods are recommended \citep{wang2016}.

This paper proposes a different route. Trend agreement between sequences is reduced to the signs of their centred increments, so each comparison is concordant or discordant. The resulting counts are analysed with likelinesses \citep{lovegrove2015}: averages of probabilities over an underlying set of distributions, which replace the elicitation of a prior density by a statement of what is known to be possible, weighted by the canonical measure on the simplex. On this object, inference is in closed form for every sample size: exact under an explicit model of exchangeable comparisons, and with a small, quantified error when centring makes the comparisons dependent (Section 4.6).

The contributions are:

\begin{enumerate}
\def\labelenumi{\arabic{enumi}.}
\tightlist
\item
  A concordance incidence grade, with its properties proved against the grey incidence axioms (Sections 3.1--3.3).
\item
  A treatment of indeterminate comparisons, whose sign falls below the resolution of the data: discarding them is exact when indeterminacy carries no information, and otherwise the grade becomes a grey number whose width measures what the data cannot decide (Section 3.4).
\item
  Closed-form inference for the grade: estimation, the likeliness that one factor outranks another, prediction of future concordance, and a baseline that corrects for searching among many candidates, including bounded underlying sets that encode partial knowledge without a prior density (Section 4).
\item
  A simulation study of estimation, prediction, search and robustness to dependence (Section 5).
\item
  Two illustrations: the worked examples of \citet[ch.~5]{liu2017} and the location of the signal--noise transition in a regularisation benchmark where the truth is known (Section 6).
\end{enumerate}

\hypertarget{literature-review}{%
\section{Literature review}\label{literature-review}}

\hypertarget{grey-incidence-models}{%
\subsection{Grey incidence models}\label{grey-incidence-models}}

Deng's model compares sequences point by point through a coefficient that depends on a distinguishing coefficient \ensuremath{\rho} and on the global minimum and maximum differences of the whole system. Liu introduced the absolute, relative and synthetic degrees, based on areas between zero-starting-point images, and later the similitude and closeness degrees, which separate the geometric shape of the curves from their proximity \citep{liu2013,liu2017}. A separate line works directly on trends: the slope incidence degree, improved by \citet{dang2004} to capture positive and negative relations and to be invariant to dimensionless transformations, and its extension to gradients of multivariate functions \citep{zhang2017}. The concordance grade of this paper belongs to this trend-based line.

\hypertarget{known-inconsistencies}{%
\subsection{Known inconsistencies}\label{known-inconsistencies}}

\citet{dong2018} compared five widely used models and showed that their incidence orders often differ and can lead to incompatible decisions, which they traced to models measuring closeness, similarity or a combination without saying so. \citet{xie2009} examined the grey incidence axioms and proved that the lower bound of almost every model is not zero and that totality and pair symmetry conflict. \citet{nowak2024} showed by simulation that the relative degree depends materially on how series are standardised: dividing by the initial value, as in the original model, gives the first terms the greatest weight, against the grey principle that the most recent data matter most, and dividing by the mean is more reliable.

\hypertarget{uncertainty-for-incidence-grades}{%
\subsection{Uncertainty for incidence grades}\label{uncertainty-for-incidence-grades}}

\citet{saavedra2013} propagated measurement error through the operations of grey relational analysis and showed that ignoring it can change the reading of a ranking. For small samples, grey confidence degrees have been combined with bootstrap resampling \citep{wang2016}. Both approaches import an assumption the grey method avoids: a noise model in the first case, and in the second the representativeness of an empirical distribution built from perhaps ten points. Neither gives the distribution of the incidence grade itself, nor a prediction of how it will evolve.

\hypertarget{likelinesses}{%
\subsection{Likelinesses}\label{likelinesses}}

\citet{lovegrove2015} defines the likeliness of an event given a histogram h as the average of its probability over an underlying set P of distributions, each weighted by f\textsuperscript{h}. No limiting process and no prior density are involved. The move follows \citet{whittle1992}, who rebuilt probability theory by taking expectation, rather than probability, as the primitive ``entity of prime interest''; Lovegrove makes the likeliness that primitive instead. This choice is what licenses the method used here: questions of how well a likeliness estimates some true probability, or of whether a sample is large, do not arise. On the full simplex S(N) he proves a Law of Succession, a Combination Theorem and an Integram Theorem, and he distinguishes likeliness from probability through Multinomial Consistency: the average of a nonlinear functional equals the functional of the average only if P is a singleton or the integram has size at most one. On S(N) the computations coincide with Bayes--Laplace under a uniform Dirichlet prior, whose predictive form was characterised by Johnson's sufficientness postulate \citep{zabell1982}. What the framework adds is the reading as an average over a set and the use of non-standard underlying sets, both essential below.

\hypertarget{theoretical-framework}{%
\section{Theoretical framework}\label{theoretical-framework}}

\hypertarget{likelinesses-on-the-simplex}{%
\subsection{Likelinesses on the simplex}\label{likelinesses-on-the-simplex}}

Let S(K) be the simplex of distributions on K outcomes and P \ensuremath{\subseteq} S(K) an underlying set. For a histogram h with \ensuremath{\omega}(h) = \ensuremath{\Sigma} h(i) observations, the likeliness of outcome i is

\[
L_P(i \mid h) = \frac{\int_P f(i)\, f^{h}\, df}{\int_P f^{h}\, df}, \qquad f^{h} = \prod_{j} f(j)^{h(j)}
\]

On P = S(K) this gives the Law of Succession, L(i \textbar{} h) = (1 + h(i))/(K + \ensuremath{\omega}(h)), and the likeliness of a future integram g follows the Integram Theorem \citep{lovegrove2015}:

\[
L_{S(K)}(g \mid h) = \frac{M(g)\,M(h)}{M(g+h)}\,\frac{|\Omega_K(\omega(h))|}{|\Omega_K(\omega(g+h))|}
\]

where M is the multinomial coefficient and \textbar \ensuremath{\Omega}\textsubscript{K}(n)\textbar{} the number of integrams of size n.

\textbf{Base measure.} The integrals over P are taken with respect to Lebesgue measure on the simplex, as in \citet{lovegrove2015}; it is the measure invariant under relabelling of the outcomes, and the one that gives the Law of Succession. Integrating against it is formally the same as averaging under a uniform density restricted to P: on S(2) the computations coincide with Bayes--Laplace (Section 2.4), and on P{[}a, b{]} with a uniform prior on {[}a, b{]}. The framework therefore does not dispense with a measure. What it avoids is the elicitation of a density shape: the analyst states which distributions are possible, and the canonical measure weights them equally. Section 4.5 reports how much the results depend on this choice.

\hypertarget{the-concordance-incidence-grade}{%
\subsection{The concordance incidence grade}\label{the-concordance-incidence-grade}}

Let X\textsubscript{0} and X\textsubscript{\textit{i}} be sequences of length n + 1 with increments \ensuremath{\Delta}x\textsubscript{\textit{j}}(k). Each increment is centred by subtracting its sequence mean, \ensuremath{\tilde\Delta}x\textsubscript{\textit{j}}(k) = \ensuremath{\Delta}x\textsubscript{\textit{j}}(k) \ensuremath{-} mean(\ensuremath{\Delta}x\textsubscript{\textit{j}}), and the comparisons are counted:

\[
c = \#\{k : \tilde\Delta x_0(k)\,\tilde\Delta x_i(k) > 0\}, \qquad d = \#\{k : \tilde\Delta x_0(k)\,\tilde\Delta x_i(k) < 0\}, \qquad r(X_0, X_i) = \frac{1 + c}{2 + c + d}
\]

The grade r is the likeliness of concordance on S(2). Centring removes a common trend, which otherwise makes almost every increment concordant: in the simulations of Section 5, raw concordance falls to the level of Deng's grade when sequences share a trend. Comparisons whose sign cannot be decided, including any with a centred increment exactly zero, are treated in Section 3.4.

\textbf{Modelling assumption.} All inference below treats the c + d comparisons of a candidate as exchangeable and extendable, so that the integram is a sufficient summary and the likelinesses of Section 3.1 apply. The computations are exact under this assumption, not for every data-generating process. Centring makes the comparisons of one candidate weakly dependent, because the centred increments of a sequence sum to zero; Section 4.6 shows that the effect is of order 1/n and how to correct it.

\hypertarget{properties-against-the-grey-incidence-axioms}{%
\subsection{Properties against the grey incidence axioms}\label{properties-against-the-grey-incidence-axioms}}

\citet[Definition 5.3.2]{liu2017} state two conditions for a degree of grey incidence: normality (0 \textless{} \ensuremath{\gamma} \ensuremath{\leq} 1, with \ensuremath{\gamma} = 1 if and only if X\textsubscript{0} = X\textsubscript{\textit{i}}) and closeness (the smaller \textbar x\textsubscript{0}(k) \ensuremath{-} x\textsubscript{\textit{i}}(k)\textbar, the larger the local coefficient). Deng's original set adds pair symmetry and totality \citep{deng1985,xie2009}. Proofs are short and follow from the definition.

\textbf{Proposition 1 (weak normality).} 1/(n\ensuremath{'} + 2) \ensuremath{\leq} r \ensuremath{\leq} (n\ensuremath{'} + 1)/(n\ensuremath{'} + 2), with n\ensuremath{'} = c + d; hence 0 \textless{} r \textless{} 1, and r = 1/2 means no evidence of relation.

\textbf{Proposition 2 (identity is not certified).} For X\textsubscript{\textit{i}} = X\textsubscript{0}, r = (n + 1)/(n + 2) \textless{} 1, and r attains this maximum for every X\textsubscript{\textit{i}} = aX\textsubscript{0} + b with a \textgreater{} 0. The strong form of normality fails on principle: a likeliness from finite data never asserts identity, and a similarity grade identifies equal trends, not equal sequences. Liu's own absolute and similitude degrees are in the same position: both equal 1 for parallel sequences \citep[Theorems 5.4.2 and 5.6.1]{liu2017}.

\textbf{Proposition 3 (symmetry).} r(X\textsubscript{0}, X\textsubscript{\textit{i}}) = r(X\textsubscript{\textit{i}}, X\textsubscript{0}) in any system, which is stronger than pair symmetry.

\textbf{Proposition 4 (totality).} r depends only on the pair, so totality fails by design. This resolves the conflict proved by \citet{xie2009}: symmetry is kept in the grade, and dependence on the system moves into inference, through the search baseline of Section 4.4.

\textbf{Proposition 5 (closeness).} r(X\textsubscript{0}, X\textsubscript{0} + M) = r(X\textsubscript{0}, X\textsubscript{0}) for any constant M, so closeness of levels fails. In the classification of \citet{dong2018}, r is a similarity model, for which the relevant property is closeness of trends.

\textbf{Proposition 6 (affine invariance and negative relation).} r(X\textsubscript{0}, aX\textsubscript{\textit{i}} + b) equals r(X\textsubscript{0}, X\textsubscript{\textit{i}}) for a \textgreater{} 0 and 1 \ensuremath{-} r(X\textsubscript{0}, X\textsubscript{\textit{i}}) for a \textless{} 0. Negative relations are represented symmetrically about 1/2, the property sought by \citet{dang2004}.

\textbf{Proposition 7 (order of comparisons).} r is invariant to time reversal and to any common permutation of the comparisons. It therefore weights all data equally, contrary to the grey principle of favouring recent information; a weighted version is possible but reintroduces a free scale for the histogram.

\hypertarget{indeterminate-comparisons}{%
\subsection{Indeterminate comparisons}\label{indeterminate-comparisons}}

A comparison is indeterminate when the sign of a centred increment cannot be decided: flat stretches, increments at the level of the noise, or interval-valued (grey) data whose consecutive intervals overlap. With three classes, concordant, indeterminate and discordant, counts (c, i, d) and n = c + i + d, the likeliness on S(3) is Dirichlet(1 + c, 1 + i, 1 + d).

\textbf{Proposition 8 (conditional invariance).} On S(3), the likeliness that a determinate comparison is concordant is (1 + c)/(2 + c + d), and \ensuremath{\theta}\textsubscript{C}/(\ensuremath{\theta}\textsubscript{C} + \ensuremath{\theta}\textsubscript{D}) is Beta(1 + c, 1 + d) whatever the value of i. Discarding indeterminate comparisons is therefore exact when indeterminacy carries no information about trend agreement, and every result of Section 4 applies to (c, d). The proof is the neutrality of the Dirichlet distribution: the ratio of two components to their sum is independent of the remaining component.

\textbf{Proposition 9 (grey concordance grade).} If indeterminacy may depend on the relation itself, the grade is only known to lie in the interval obtained by resolving every indeterminate comparison as discordant or as concordant:

\[
r^{-} = \frac{1+c}{2+n} \;\le\; \frac{1+c}{2+c+d} \;\le\; r^{+} = \frac{1+c+i}{2+n}
\]

The interval contains the conditional grade, and its width, i/(2 + n), is the greyness that the data cannot remove. Ordering likelinesses are bounded in the same way, by resolving the indeterminate comparisons of the two candidates in opposite directions.

Indeterminacy needs a threshold, and a default that involves no choice is the resolution of the data. Each published value x stands for the grey number {[}x \ensuremath{-} u/2, x + u/2{]}, u being the unit of its last digit. Centred increments are linear in the data, so each is known to within (u/2)\ensuremath{\Sigma}\textbar w\textbar, w being its coefficients, and a comparison is indeterminate when either interval contains zero. A centred increment that is exactly zero is indeterminate under any threshold.

\hypertarget{inference-for-the-concordance-grade}{%
\section{Inference for the concordance grade}\label{inference-for-the-concordance-grade}}

Throughout, a candidate has c concordant and d discordant comparisons, n = c + d, and its likeliness of concordance \ensuremath{\theta} is Beta(1 + c, 1 + d) on S(2).

\hypertarget{estimation}{%
\subsection{Estimation}\label{estimation}}

The grade r = (1 + c)/(2 + n) lies between 1/2 and c/n \citep[Theorem 5]{lovegrove2015}. Its mean squared error, {[}n\ensuremath{\theta}(1 \ensuremath{-} \ensuremath{\theta}) + (1 \ensuremath{-} 2\ensuremath{\theta})\textsuperscript{2}{]}/(n + 2)\textsuperscript{2}, is smaller than that of the relative frequency, \ensuremath{\theta}(1 \ensuremath{-} \ensuremath{\theta})/n, for \ensuremath{\theta} in roughly {[}0.14, 0.86{]} at every n; averaged over \ensuremath{\theta}, the reduction is 29\% at n = 5 and 4\% at n = 50.

\hypertarget{ordering-two-candidates}{%
\subsection{Ordering two candidates}\label{ordering-two-candidates}}

\textbf{Proposition 10.} For independent candidates with integer parameters a\textsubscript{\textit{j}} = 1 + c\textsubscript{\textit{j}} and b\textsubscript{\textit{j}} = 1 + d\textsubscript{\textit{j}},

\[
L(\theta_2 > \theta_1) = \sum_{i=0}^{a_2-1} \frac{B(a_1+i,\; b_1+b_2)}{(b_2+i)\,B(1+i,\; b_2)\,B(a_1,\; b_1)}
\]

and L(\ensuremath{\theta}\textsubscript{1} \textgreater{} \ensuremath{\theta}\textsubscript{2}) = 1 \ensuremath{-} L(\ensuremath{\theta}\textsubscript{2} \textgreater{} \ensuremath{\theta}\textsubscript{1}). For c\textsubscript{1} = 9 of 10 against c\textsubscript{2} = 7 of 10, L(\ensuremath{\theta}\textsubscript{1} \textgreater{} \ensuremath{\theta}\textsubscript{2}) = 0.845; for 6 of 10 against 5 of 10 it is 0.665, an ordering the plug-in grade reports with certainty. Candidates compared against a common reference share its signs and are not independent; Section 7 discusses this limitation.

\hypertarget{prediction}{%
\subsection{Prediction}\label{prediction}}

The likeliness of k concordances in m future comparisons is given by the Integram Theorem and is beta-binomial.

\textbf{Proposition 11.} The plug-in predictive Binomial(m, r) has the same mean as the Integram Theorem but a smaller variance:

\[
\frac{\operatorname{Var}_{\text{Integram}}}{\operatorname{Var}_{\text{plug-in}}} = \frac{n + 2 + m}{n + 3} \;\longrightarrow\; 1 + \frac{m}{n} \quad (n \to \infty,\; m/n \text{ fixed})
\]

\emph{Proof.} The beta-binomial variance with \ensuremath{\alpha} + \ensuremath{\beta} = n + 2 is m r(1 \ensuremath{-} r)(\ensuremath{\alpha} + \ensuremath{\beta} + m)/(\ensuremath{\alpha} + \ensuremath{\beta} + 1). \hfill\ensuremath{\square}

The error of the plug-in therefore depends on the horizon relative to the sample, not on the sample size: more data do not repair it. This quantifies Lovegrove's Multinomial Consistency, by which the plug-in is valid only for a singleton set or an integram of size at most one.

\hypertarget{a-baseline-for-searches}{%
\subsection{A baseline for searches}\label{a-baseline-for-searches}}

Grey analysis typically selects the best of K candidates. With no data and P = S(2), every concordance level 0, \ldots, n is equally likely (Combination Theorem), so L(C \ensuremath{\geq} c) = (n \ensuremath{-} c + 1)/(n + 1). A pattern is declared when the likeliness that the maximum of K unrelated candidates reaches the observed c\textsubscript{max} is small:

\[
L_{\text{base}}(\max \ge c_{\max}) = 1 - F_{\text{base}}(c_{\max} - 1)^K \le 0.05
\]

The choice of baseline is decisive (Section 5.3): the fair coin assumes more than is known, while S(2) assumes less.

\hypertarget{bounded-underlying-sets}{%
\subsection{Bounded underlying sets}\label{bounded-underlying-sets}}

Partial knowledge, such as knowing that concordance between unrelated series cannot be extreme, is encoded as P{[}a, b{]} = \{f \ensuremath{\in} S(2) : f(C) \ensuremath{\in} {[}a, b{]}\}. With Z(\ensuremath{\alpha}, \ensuremath{\beta}) = \ensuremath{\int_a^b} t\textsuperscript{\ensuremath{\alpha}\ensuremath{-}1}(1 \ensuremath{-} t)\textsuperscript{\ensuremath{\beta}\ensuremath{-}1} dt, a difference of incomplete beta functions:

\[
L_{P[a,b]}(\text{C} \mid h) = \frac{Z(2+c,\, 1+d)}{Z(1+c,\, 1+d)}, \qquad L_{P[a,b]}(k \mid h, m) = \binom{m}{k}\frac{Z(1+c+k,\, 1+d+m-k)}{Z(1+c,\, 1+d)}
\]

On P{[}0.3, 0.7{]}, 9 concordances out of 10 give a likeliness of 0.627 instead of 0.833, and the no-data likeliness of c \ensuremath{\geq} 9 falls from 0.18 to 0.030. No parametric family is chosen at any point. The choice of base measure matters most on S(2): replacing Lebesgue measure by Beta(\ensuremath{\alpha}, \ensuremath{\alpha}) with \ensuremath{\alpha} \ensuremath{\in} \{0.5, 2\} changes L(\ensuremath{\theta}\textsubscript{1} \textgreater{} \ensuremath{\theta}\textsubscript{2}) for 9 of 10 against 7 of 10 from 0.845 to between 0.81 and 0.87, and the no-data likeliness of c \ensuremath{\geq} 9 in 10 comparisons from 0.18 to between 0.11 and 0.27, so the S(2) search baseline is sensitive to it. On P{[}0.3, 0.7{]} the same ordering changes by less than 0.005 and the tail stays between 0.029 and 0.031: once the set is bounded, the set rather than the measure carries the assumption.

\hypertarget{dependent-comparisons}{%
\subsection{Dependent comparisons}\label{dependent-comparisons}}

When concordance signs have lag-one autocorrelation \ensuremath{\rho}, an approximate correction rescales the integram to its effective size, h\textsubscript{ef} = h(1 \ensuremath{-} \ensuremath{\rho})/(1 + \ensuremath{\rho}), with \ensuremath{\rho} estimated from the sign sequence itself. The factor is the large-sample variance inflation of the mean of a stationary two-state Markov chain, so the correction matches the first two moments of the concordance proportion asymptotically and nothing more. It keeps the properties of Section 3.3, because both counts are rescaled by the same factor, and it changes the point grade only slightly. It does not make the likelinesses exact: it ignores the finite-sample and higher-order structure of the chain, and Section 5.4 shows that it is not sufficient for prediction. The scale of the histogram is then set by the data rather than chosen, but it is estimated, and with fewer than about 20 comparisons the estimate is unreliable.

A second source of dependence is the centring itself. The centred increments of a sequence sum to zero, so each sequence has at least one positive and one negative centred increment. This does not restrict the count, since identical sign patterns give c = n and opposite ones d = n, but it does correlate the comparisons.

\textbf{Proposition 12 (dependence induced by centring).} Let X\textsubscript{0} and X\textsubscript{\textit{i}} be independent, each with i.i.d. increments from a continuous symmetric distribution. Then the centred comparisons are exchangeable but not independent: any two of them have covariance \ensuremath{\kappa}\textsubscript{0}\ensuremath{\kappa}\textsubscript{\textit{i}}, where \ensuremath{\kappa}\textsubscript{\textit{j}} is the correlation between the signs of two centred increments of X\textsubscript{\textit{j}}, and Var(c) = (n/4){[}1 + (n \ensuremath{-} 1)\ensuremath{\kappa}\textsubscript{0}\ensuremath{\kappa}\textsubscript{\textit{i}}{]}. For Gaussian increments \ensuremath{\kappa}\textsubscript{0} = \ensuremath{\kappa}\textsubscript{\textit{i}} = \ensuremath{-}(2/\ensuremath{\pi}) arcsin(1/(n \ensuremath{-} 1)), so the excess variance is of order 1/n.

\emph{Proof.} Write s\textsubscript{\textit{k}} = \ensuremath{\sigma}\textsubscript{\textit{k}}\ensuremath{\tau}\textsubscript{\textit{k}} for the product of the centred signs of the two sequences. The sign vectors \ensuremath{\sigma} and \ensuremath{\tau} are independent and each exchangeable, so the products are exchangeable, E{[}s\textsubscript{\textit{k}}{]} = 0 and Cov(s\textsubscript{\textit{k}}, s\textsubscript{\textit{l}}) = E{[}\ensuremath{\sigma}\textsubscript{\textit{k}}\ensuremath{\sigma}\textsubscript{\textit{l}}{]} E{[}\ensuremath{\tau}\textsubscript{\textit{k}}\ensuremath{\tau}\textsubscript{\textit{l}}{]}. Two centred Gaussian increments of the same sequence have correlation \ensuremath{-}1/(n \ensuremath{-} 1), and the signs of a Gaussian pair with correlation r have correlation (2/\ensuremath{\pi}) arcsin r. \hfill\ensuremath{\square}

The factor 1 + (n \ensuremath{-} 1)\ensuremath{\kappa}\textsubscript{0}\ensuremath{\kappa}\textsubscript{\textit{i}} is 1.14 at n = 4, 1.10 at n = 5 and 1.05 at n = 10. It can be absorbed approximately by the same device as above, with effective size n/{[}1 + (n \ensuremath{-} 1)\ensuremath{\kappa}\textsubscript{0}\ensuremath{\kappa}\textsubscript{\textit{i}}{]}. For Gaussian increments centring leaves the estimand unchanged: each centred pair keeps the correlation \ensuremath{\rho} of the original increments, so \ensuremath{\theta} = 1/2 + arcsin(\ensuremath{\rho})/\ensuremath{\pi} with or without centring.

\hypertarget{simulation-study}{%
\section{Simulation study}\label{simulation-study}}

Five questions were examined: whether ordering likelinesses are calibrated and how much power the grade has (5.1), whether predictions cover (5.2), whether searches produce spurious patterns (5.3), how dependence between comparisons affects all of these (5.4), and what happens when the underlying set is wrong (5.5).

\hypertarget{calibration-and-power-of-the-ordering}{%
\subsection{Calibration and power of the ordering}\label{calibration-and-power-of-the-ordering}}

For 40,000 pairs of candidates, L(\ensuremath{\theta}\textsubscript{1} \textgreater{} \ensuremath{\theta}\textsubscript{2}) was compared with the observed frequency of \ensuremath{\theta}\textsubscript{1} \textgreater{} \ensuremath{\theta}\textsubscript{2}. When the true concordances were drawn uniformly, the expected calibration error was below 0.005. When they were concentrated in {[}0.3, 0.7{]}, S(2) was overconfident (0.099 at n = 5, 0.039 at n = 20), and the bounded set P{[}0.3, 0.7{]} reduced the error by a factor of four to five (0.023 and 0.009).

Power was measured with two candidates against one reference, one with increment correlation \ensuremath{\rho}\textsubscript{1} and one unrelated, with n = 10 and n = 20 and 3,000 replicates per setting. Centred concordance ordered them correctly more often than Deng's grade, by 7 to 20 points, and Liu's absolute degree, by 7 to 26 points. Without centring, a common trend reduced concordance to Deng's level. Correlation of increments was more powerful still, by 2 to 11 points, because it uses magnitudes and is optimal under Gaussian noise; but its small-sample inference requires that assumption.

\hypertarget{prediction-coverage}{%
\subsection{Prediction coverage}\label{prediction-coverage}}

For \ensuremath{\theta} \ensuremath{\in} \{0.5, 0.7, 0.9\}, n \ensuremath{\in} \{5, 10, 20, 40\} and horizons m \ensuremath{\in} \{n/2, n, 2n\}, the exact coverage of nominal 90\% predictive sets was computed by enumeration.

\begin{longtable}[]{@{}lll@{}}
\toprule\noalign{}
Horizon & Integram Theorem & Plug-in binomial \\
\midrule\noalign{}
\endhead
\bottomrule\noalign{}
\endlastfoot
m = n/2 & 0.93--1.00 & 0.88--0.99 \\
m = n & 0.91--0.99 & 0.78--0.99 \\
m = 2n & 0.92--0.97 & 0.70--0.82 \\
\end{longtable}

The plug-in loses coverage as the horizon grows, and at m = 2n its divergence from the Integram distribution grows with n (0.22 at n = 5, 0.41 at n = 40), as Proposition 11 predicts.

\hypertarget{spurious-patterns-in-searches}{%
\subsection{Spurious patterns in searches}\label{spurious-patterns-in-searches}}

Among K candidates with n comparisons, unrelated candidates had \ensuremath{\theta} uniform in {[}0.5 \ensuremath{-} s, 0.5 + s{]}, since unrelated series need not concord exactly half the time. The best candidate was declared a pattern by the rule of Section 4.4 (20,000 replicates per setting).

\begin{longtable}[]{@{}
  >{\raggedright\arraybackslash}p{(\columnwidth - 8\tabcolsep) * \real{0.2000}}
  >{\raggedright\arraybackslash}p{(\columnwidth - 8\tabcolsep) * \real{0.2000}}
  >{\raggedright\arraybackslash}p{(\columnwidth - 8\tabcolsep) * \real{0.2000}}
  >{\raggedright\arraybackslash}p{(\columnwidth - 8\tabcolsep) * \real{0.2000}}
  >{\raggedright\arraybackslash}p{(\columnwidth - 8\tabcolsep) * \real{0.2000}}@{}}
\toprule\noalign{}
\begin{minipage}[b]{\linewidth}\raggedright
Unrelated \ensuremath{\theta} range (n = 20, K = 20)
\end{minipage} & \begin{minipage}[b]{\linewidth}\raggedright
Fair coin
\end{minipage} & \begin{minipage}[b]{\linewidth}\raggedright
P{[}0.3, 0.7{]}
\end{minipage} & \begin{minipage}[b]{\linewidth}\raggedright
P{[}0.2, 0.8{]}
\end{minipage} & \begin{minipage}[b]{\linewidth}\raggedright
S(2)
\end{minipage} \\
\midrule\noalign{}
\endhead
\bottomrule\noalign{}
\endlastfoot
0.5 & 2.7\% & 0.04\% & 0.0\% & 0\% \\
0.3--0.7 & 24.9\% & 1.6\% & 0.1\% & 0\% \\
0.2--0.8 & 62.6\% & 10.2\% & 1.3\% & 0\% \\
Power, one candidate with \ensuremath{\theta} = 0.9 & 70--86\% & 38--40\% & 12\% & 0\% \\
\end{longtable}

The fair coin is calibrated only when unrelated candidates concord exactly half the time; with K = 100 its spurious rate reached 86\%. S(2) never detects anything, true patterns included. A bounded set covering the real range of unrelated concordance keeps spurious patterns below 5\% while keeping intermediate power; one that is too narrow rises to 10\%. Unrelated series with their own drifts produced raw concordances between 0.3 and 0.7 and 23\% spurious patterns under the fair coin; centring lowered this to 3.3\%.

\hypertarget{dependent-comparisons-1}{%
\subsection{Dependent comparisons}\label{dependent-comparisons-1}}

Concordance signs were generated as a two-state Markov chain with autocorrelation \ensuremath{\rho} \ensuremath{\in} \{0, 0.2, 0.4, 0.6\} (20,000 replicates; 40,000 pairs for ordering).

\begin{itemize}
\tightlist
\item
  Without correction, coverage of the 90\% interval for \ensuremath{\theta} fell by 7 to 8 points already at \ensuremath{\rho} = 0.2, and more at larger n (0.80 at n = 40), because more comparisons narrow the interval while dependence does not dilute.
\item
  The effective-size correction with known \ensuremath{\rho} kept coverage between 0.91 and 0.96. With \ensuremath{\rho} estimated from the data it recovered most of the loss for n \ensuremath{\geq} 20 (0.86 at n = 40 and \ensuremath{\rho} = 0.6), but not at n = 10, where the autocorrelation of ten binary signs cannot be estimated.
\item
  The correction reduced the ordering calibration error to between a third and a half. For prediction it was not enough: even with known \ensuremath{\rho}, coverage stayed between 0.83 and 0.91 for \ensuremath{\rho} \ensuremath{\geq} 0.4.
\item
  Increments following AR(1) processes with coefficient up to 0.9 induced sign autocorrelation of about 0.2 after centring, the regime where the correction works.
\end{itemize}

Dependence from centring was checked with Gaussian increments (20,000 replicates, increment correlation \ensuremath{\rho} \ensuremath{\in} \{0, 0.5, 0.8\}). The variance of c exceeded the binomial by 7--16\% at n = 4, in agreement with Proposition 12, and by at most 5\% for n \ensuremath{\geq} 10. Coverage of the 90\% interval for \ensuremath{\theta} fell by up to 5 points at n = 4, 2.4 points at n = 5 and at most 1 point for n \ensuremath{\geq} 10, while predictive coverage was unchanged. The textbook examples of Section 6.1, with four to six increments, lie where the effect is largest; their ordering likelinesses, already weak, are slightly overconfident.

\hypertarget{misspecified-underlying-sets}{%
\subsection{Misspecified underlying sets}\label{misspecified-underlying-sets}}

The calibration results of Sections 5.1 and 5.3 were obtained with concordances generated inside the chosen set, so they show internal consistency rather than calibration in applications. To see what happens when the set is wrong, true concordances were drawn from U{[}0.2, 0.8{]}, U{[}0, 1{]} and Beta(5, 2), a population biased towards concordance, as well as from U{[}0.3, 0.7{]} (6,000 pairs; n = 20, with similar results at n = 10). Each cell gives the calibration error of L(\ensuremath{\theta}\textsubscript{1} \textgreater{} \ensuremath{\theta}\textsubscript{2}) and the coverage of the 90\% interval for \ensuremath{\theta}.

\begin{longtable}[]{@{}llll@{}}
\toprule\noalign{}
True concordances & S(2) & P{[}0.3, 0.7{]} & P{[}0.2, 0.8{]} \\
\midrule\noalign{}
\endhead
\bottomrule\noalign{}
\endlastfoot
U{[}0.3, 0.7{]} & 0.045 / 0.91 & 0.008 / 0.90 & 0.034 / 0.93 \\
U{[}0.2, 0.8{]} & 0.016 / 0.90 & 0.035 / 0.60 & 0.007 / 0.90 \\
U{[}0, 1{]} & 0.008 / 0.89 & 0.066 / 0.37 & 0.033 / 0.55 \\
Beta(5, 2) & 0.015 / 0.90 & 0.090 / 0.37 & 0.046 / 0.58 \\
\end{longtable}

A bounded set is well calibrated when it contains the truth and fails badly when it does not: P{[}0.3, 0.7{]} covered \ensuremath{\theta} in only 36--60\% of cases drawn outside its range, and its ordering error rose to between 0.04 and 0.09, worse than that of S(2). A wider set degraded less, and S(2), which excludes nothing, was never far from calibration. Bounded sets are therefore justified as baselines for unrelated candidates, whose concordance can be bounded from the drift and noise of the series before the comparisons are made, and not as models for the concordance of related candidates, which can be arbitrarily high. The benchmark of Section 6.2, where concordance is produced by a physical mechanism rather than drawn from a set, is the only out-of-model check in this paper.

\hypertarget{illustrative-examples}{%
\section{Illustrative examples}\label{illustrative-examples}}

\hypertarget{worked-examples-of-liu-et-al.-2017-ch.-5}{%
\subsection{\texorpdfstring{Worked examples of \citet[ch.~5]{liu2017}}{Worked examples of Liu et al. (2017, ch.~5)}}\label{worked-examples-of-liu-et-al.-2017-ch.-5}}

Three worked examples of the chapter have several candidates and a published ordering. Others were not used: Example 5.4.1 has a single candidate, and the lag analysis of Example 5.8.1, a natural search over 16 indices and 25 lags, publishes the grades but not the monthly series. All likelinesses below use centred increments; the textbook sequences have four to six increments.

\textbf{Example 5.3.1 (Deng's grade).} Chinese GDP is the reference, and the value added by the primary, secondary and tertiary industries are the candidates. Deng's grade ranks them tertiary (0.855) \textgreater{} secondary (0.788) \textgreater{} primary (0.566). All three concord with GDP in 4 of 4 comparisons, centred or not, so r = 0.833 for each and every pairwise ordering has likeliness exactly 0.5. Because the three integrams are identical, this conclusion does not depend on the independence assumption of Proposition 10. Deng's ranking therefore comes from levels, not trends: Deng's grade measures closeness and the concordance grade similarity, and the example shows within the textbook how the two answer different questions \citep{dong2018}.

\textbf{Example 5.7.1 (superiority analysis).} Three characteristic sequences are each compared with five factors through the absolute degree, giving a 3 \ensuremath{\times} 5 matrix of grades between 0.50 and 0.91. With four increments, concordance takes only the values 2/4, 3/4 and occasionally 1/4. The largest contrast in the matrix, 3/4 against 1/4 for the first characteristic, gives an ordering likeliness of 0.90; a typical contrast of 3/4 against 2/4 gives 0.74. As a search, even a perfect 4/4 among five factors would not be exceptional: the likeliness that the maximum of five unrelated factors reaches 4 is 0.85 under the fair coin and 0.92 under S(2). With these data the matrix cannot separate the factors, whatever the grades suggest.

\textbf{Example 5.6.1 (similitude against closeness).} The reference X\textsubscript{1} fluctuates; X\textsubscript{2} is observed at five of its seven times and X\textsubscript{3} at three, and the textbook fills the gaps by linear interpolation. It finds X\textsubscript{2} more similar to X\textsubscript{1} (similitude 0.91 against 0.83) but X\textsubscript{3} closer (closeness 0.73 against 0.34). Concordance is computed only between observed times, because interpolated increments are deterministic and would add comparisons that were never observed. The data have two decimals, and this matters: the third centred increment of X\textsubscript{1} is exactly zero (0.01 \ensuremath{-} 0.01), and the fourth of X\textsubscript{2}, 0.0075 \ensuremath{\pm} 0.01, has no determinate sign at that resolution. With the indeterminate class of Section 3.4, X\textsubscript{2} has (c, i, d) = (2, 2, 0), so r = 0.75 with grey grade {[}0.50, 0.83{]}, and X\textsubscript{3} is discordant in both of its two comparisons, r = 0.25 with no greyness. Then L(\ensuremath{\theta}\textsubscript{2} \textgreater{} \ensuremath{\theta}\textsubscript{3}) = 0.95, within {[}0.82, 0.98{]}, in agreement with the similitude degree, as expected of a similarity grade; but half of the evidence for X\textsubscript{2} lies below the resolution of the data. Treating the zero increment as positive, as floating-point arithmetic may do, would report 4 of 4 and 0.98. With the interpolated sequences, X\textsubscript{2} gives (3, 2, 0) and X\textsubscript{3} (2, 0, 3), and the ordering likeliness drops to 0.93, within {[}0.72, 0.97{]}: interpolation manufactures evidence, and the concordance grade makes this visible. In Examples 5.3.1 and 5.7.1 every centred increment exceeds the resolution, and the grey grades have zero width.

\hypertarget{the-signalnoise-transition-in-a-regularisation-benchmark}{%
\subsection{The signal--noise transition in a regularisation benchmark}\label{the-signalnoise-transition-in-a-regularisation-benchmark}}

The second example has a known truth. It uses a replication of the benchmark of \citet{bauer2011} in the spectral domain: 100 operators \ensuremath{\times} 5 solutions \ensuremath{\times} 5 data sets, giving 2,500 instances of 300 coefficients with coloured Gaussian noise. Singular values \ensuremath{\sigma}\textsubscript{\textit{i}} decrease; the observed coefficients are b\textsubscript{\textit{i}} = \ensuremath{\sigma}\textsubscript{\textit{i}}\ensuremath{\xi}\textsubscript{\textit{i}} + e\textsubscript{\textit{i}}, and the noise e\textsubscript{\textit{i}} is known. In the Picard plot, \textbar b\textsubscript{\textit{i}}\textbar{} decays with \ensuremath{\sigma}\textsubscript{\textit{i}} while signal dominates and stops following it once noise does. The true transition i* is the first index where the expected signal magnitude falls below the noise level; 1,825 instances have an interior transition, with median i* = 7.

Each step contributes a concordance indicator, 1 if \textbar b\textsubscript{\textit{i}}\textbar{} decreases. A change point m splits the indicators: before m, concordance lies in a signal set P\textsubscript{signal}; after m, it is noise with \ensuremath{\theta} = 1/2 exactly. The likeliness of the split is

\[
L(m \mid \text{data}) \;\propto\; \frac{1}{|P_{\text{signal}}|}\int_{P_{\text{signal}}} \theta^{c_m}(1-\theta)^{m-c_m}\,d\theta \;\cdot\; 2^{-(N-m)}
\]

whose mode gives î and whose distribution gives a 90\% likeliness interval for i*.

\begin{longtable}[]{@{}
  >{\raggedright\arraybackslash}p{(\columnwidth - 4\tabcolsep) * \real{0.3333}}
  >{\raggedright\arraybackslash}p{(\columnwidth - 4\tabcolsep) * \real{0.3333}}
  >{\raggedright\arraybackslash}p{(\columnwidth - 4\tabcolsep) * \real{0.3333}}@{}}
\toprule\noalign{}
\begin{minipage}[b]{\linewidth}\raggedright
Method
\end{minipage} & \begin{minipage}[b]{\linewidth}\raggedright
Within factor 2 of i*
\end{minipage} & \begin{minipage}[b]{\linewidth}\raggedright
Coverage of the 90\% interval
\end{minipage} \\
\midrule\noalign{}
\endhead
\bottomrule\noalign{}
\endlastfoot
Concordance, consecutive indices, P\textsubscript{signal} = P{[}0.5, 1{]} & 75.5\% & 0.914 \\
Concordance, consecutive indices, P\textsubscript{signal} = S(2) & 73.6\% & 0.845 \\
Concordance, geometric grid (ratio 1.25), P{[}0.5, 1{]} & 80.5\% & 0.999 \\
Magnitude heuristic (moving average below 1.5 \ensuremath{\times} tail median) & 48.3\% & --- \\
\end{longtable}

The bounded set gives calibrated intervals and S(2) is overconfident, the same pattern as in Section 5.1. Intervals are wide, with a median ratio of 54 to 64 between their ends, because concordance declines gradually through a mixed zone after the transition that a two-regime model does not represent. Used to choose the Tikhonov parameter as \ensuremath{\lambda} = \ensuremath{\sigma}\textsuperscript{2}\textsubscript{î}, the detector reaches 31--37\% acceptance with \ensuremath{\eta} \ensuremath{\leq} 1.5 on test cycles, mid-table among standard rules, and even the exact transition reaches only 75\%. Its value is as a density-free observable that needs no true solution, not as a parameter-choice rule.

\hypertarget{discussion}{%
\section{Discussion}\label{discussion}}

\hypertarget{what-the-underlying-set-decides}{%
\subsection{What the underlying set decides}\label{what-the-underlying-set-decides}}

The most consequential choice in the method is not the grade but the underlying set. In searches, the fair coin assumes more than is known and S(2) less; only a bounded set that covers the plausible range of unrelated concordance controlled spurious patterns while keeping power. The same set calibrated ordering likelinesses and transition intervals. This is precisely the situation that motivates likelinesses: the analyst rarely knows the form of a prior density but often knows its range. The range must be fixed before the matched comparisons are seen, and Section 5.5 shows the cost of getting it wrong. One procedure, not evaluated here, is to take {[}a, b{]} from the concordances of mismatched pairs, the reference against candidates shifted in time or taken from other systems, which share the drift and noise of the data but not their relation. Whether such a range keeps the search baseline calibrated is the main open question of the method.

\hypertarget{limitations}{%
\subsection{Limitations}\label{limitations}}

\begin{itemize}
\tightlist
\item
  \textbf{Common reference.} Candidates compared against the same reference share its signs, so the closed-form ordering of Proposition 10 assumes an independence that does not hold. A joint treatment, or a bound on the error of the approximation, is needed.
\item
  \textbf{Magnitudes.} Reducing comparisons to signs costs power against increment correlation. The exchange is deliberate: closed-form inference without a noise model, in return for less power.
\item
  \textbf{Dependence.} With autocorrelated signs, the effective-size correction is reliable for estimation and ordering when n \ensuremath{\geq} 20. Below that, the autocorrelation must be fixed as an assumption, and for prediction under strong dependence the sign process itself has to be modelled. Centring adds a weak dependence within each candidate, of order 1/n (Proposition 12), which matters only for the shortest sequences.
\item
  \textbf{Recency.} The grade weights all comparisons equally; weighting reintroduces a free histogram scale.
\item
  \textbf{Indeterminate comparisons.} Section 3.4 replaces the choice of a threshold by the resolution of the data, and the grey grade reports how much of a conclusion rests below it. When indeterminacy may depend on the relation, the interval is conservative; a model of that dependence would narrow it, and a larger threshold, such as a noise level, widens it in a way that can itself be reported.
\end{itemize}

\textbf{An experimental illustration.} Three of these limitations appear together in laser-heating experiments on an epoxy (ESO--CA) nanocomposite with embedded gold nanoparticles, of the type described by \citet{otero2022}. Their purpose is thermal characterisation: the temperature rise produced as the nanoparticles convert laser light into heat is matched with a finite-element model to estimate the thermal properties of the composite, so a trend-agreement grade would serve to check whether the simulated heating curves follow the measured ones. Two nanoparticle concentrations were heated at six and seven source powers between 100 and 700 mW, with temperature recorded every second at 0.25 \textdegree{}C resolution; the laser was switched off after about 340 s. First, raw concordance between curves at different powers looked almost complete (conditional grade 0.97--0.99 at 1 s sampling), but 81--91\% of the comparisons fell below the 0.25 \textdegree{}C resolution, and the grey grade of Section 3.4 spanned {[}0.09--0.19, 1.00{]}. Coarser steps trade comparisons for determinacy: at 30 s the grade was 0.92--0.94 within {[}0.71--0.76, 0.95{]}, so the greyness offers a criterion for the time scale of the comparison. Even then, concordance reflects the switching protocol, which fixes the signs: any model with the same on and off times would concord equally, so concordance cannot validate the time constant of a monotone response. Second, the response per watt fell with power, from 193 to 159 K/W at the higher concentration, a sublinearity that lives in the level and that concordance ignores by construction; model validation needs a closeness measure alongside it. Third, residuals of every curve taken against the lowest-power curve concorded in 82\% of comparisons, which suggested a shared nonlinear pattern. The grey grade flags it as undecidable on its own: at 10 s sampling, 89--94\% of the residual comparisons are indeterminate, and the grade spans {[}0.09, 0.97{]} and {[}0.14, 0.95{]} for the two concentrations. A smoothed reference confirms it: the figure fell to 0.50--0.53, within a permutation null (95th percentile 0.63--0.65), so the apparent pattern was the noise of the common reference. Here indeterminacy concentrates in plateaus and slow cooling, so it depends on the phase of the experiment; the conditional grade of Proposition 8 would mislead, and the interval of Proposition 9 is the one to report.

\hypertarget{implications}{%
\subsection{Implications}\label{implications}}

For research on grey incidence, the concordance grade shows that there is no need to solve the totality--symmetry conflict inside a grade: system dependence can live in inference instead.

\hypertarget{conclusions}{%
\section{Conclusions}\label{conclusions}}

Grey incidence can have a closed-form sampling theory without giving up its refusal to assume probability densities. Reducing trend agreement to the signs of centred increments turns each comparison into an observation on S(2), where likelinesses give closed forms for estimation, ordering, prediction and search. The resulting grade is symmetric in any system and moves system dependence from the grade into inference, which resolves a conflict between the grey incidence axioms. Its predictions correct a plug-in error that does not shrink with sample size; bounded underlying sets turn partial knowledge into statements that are calibrated when that knowledge is right, and control spurious patterns in searches; and in a benchmark with known truth, its intervals cover at the nominal level. Future work concerns a joint treatment of candidates that share a reference, criteria for choosing bounded sets before seeing the data, and applications to inverse problems, where comparing the trends of reconstructions with the truth is a natural use of the grade.

\hypertarget{acknowledgements}{%
\section*{Acknowledgements}\label{acknowledgements}}
\addcontentsline{toc}{section}{Acknowledgements}

The author thanks Dr Frank Bauer for generously sharing the code of his regularisation benchmark, on which the replication used in Section 6.2 is based.

\bibliographystyle{plainnat}
\bibliography{concordance_grey_incidence}

\end{document}